\documentclass{acm_proc_article-sp}

\newtheorem{theorem}{Theorem}
\newtheorem{definition}{Definition}
\newtheorem{lemma}{Lemma}

\newcommand{\rrarrow}{\longrightarrow}

\newcommand{\catarrow}[2]{\ \mathrel{\mathop{\kern 0pt \rrarrow} \limits_{#1}^{#2}}\ }
\newcommand{\Catarrow}[2]{\mathrel{\mathop{\kern 0pt \Longrightarrow} \limits_{#1}^{#2}}}

\long\def\comment#1{}

\begin{document}

\title{On the generalized dining philosophers problem\titlenote{Work supported by the 
NSF-POWRE grant EIA-0074909.}}
\numberofauthors{1}
\author{
\alignauthor{Oltea Mihaela Herescu \qquad \qquad Catuscia Palamidessi}\\ \ 
\\
\affaddr{Department of Computer Science and Engineering,
The Pennsylvania State University\\
Pond Lab, University Park, PA 16802-6106  USA}\\
\email{\{herescu,catuscia\}@cse.psu.edu}
}
\date{}
\maketitle

\begin{abstract}
We consider a generalization of the dining 
philosophers problem to arbitrary connection topologies. We
focus on symmetric, fully distributed systems, and we address 
the problem of guaranteeing progress and lockout-freedom, 
even in presence of adversary schedulers, by
using randomized algorithms. We show that the well-known
al\-go\-rithms of Lehmann and Rabin do not work in the generalized
case, and we propose an alternative algorithm based on the idea of
letting the philosophers assign a random priority to their
adjacent forks.
\end{abstract}


\section{Introduction}

The problem of the dining philosophers, proposed by Dijkstra in 
\cite{Dijkstra:71:AI}, is a
very popular example of control problem in distributed systems, and has
become a typical benchmark for testing the expressiveness of concurrent languages
and of resource allocation strategies.

The typical dining philosophers sit at a round table
in positions  alternated with forks, so that there is a fork between each two
philosophers, and a philosopher  between each two forks. Each philosopher can 
pick up only the forks immediately to his right and to his left, one at the time, 
and needs both of them to eat.
The aim is to make sure that if there are hungry philosophers 
then some of them  will eventually eat ({\it progress}), or,
more ambitiously, that every hungry philosopher
 will eventually eat ({\it lockout-freedom}).

The solutions to the problem of the dining philosophers depend
fundamentally on the assumptions made on the system. If we do not
impose an initial symmetry, or do not impose that the system be
completely distributed, then several solutions are possible. Some
examples are:
\begin{itemize}
\item The forks are ordered and each philosopher tries to
get first the adjacent fork which is higher in the ordering.
\item The philosophers are colored yellow and blue alternateadly. 
The yellow philosophers try to get first the fork to their left. 
The blue ones try to get first the fork to their right.
\item There is a central monitor which controls the assignment of 
the forks to the philosophers.
\item There is a box with $n-1$ tickets, where $n$ is the number of the 
philosophers, and each philosopher must get a ticket before trying 
to get the forks.
\end{itemize}

In the first two solutions above, 
the system is not symmetric. In the last two, it is not fully 
distributed.

Of course, the problem becomes much more challenging when  we
impose the conditions of symmetry and full distribution. More
precisely, symmetry means that the philosophers are
indistinguishable, as well as the forks. The philosophers run the
same program, and both the  forks and the philosophers are all in the same
initial state. Full distribution means that there are no other
processes except the philosophers, there is no central memory, all
philosophers run independently, and the only possible interaction
is via a shared fork.

The conditions of symmetry and full distribution are interesting 
also for practical considerations: in several cases it is desirable
to consider systems which are made of copies of the same
components, and have no central control or shared memory. 
In particular, symmetry offers advantages at the level 
of reasoning about the system, as it allows a greater 
modularity, and at the level of implementation of concurrent 
languages, as it allows a compositional compilation. 
Full distribution is usually convenient as it avoids 
the overhead of a centralized control. 

Lehmann and Rabin have shown in \cite{Rabin:94:HOARE}
the remarkable result that there are no {\it deterministic}
solutions to the dining philosophers problem, 
if symmetry and full distributions are imposed, and if 
no assumption (except fairness) are made on the scheduler.
The only possible solution, in such conditions, 
are {\it randomized} algorithms, that allow to
eventually break the initial symmetry with probability $1$. 
In \cite{Rabin:94:HOARE} two such algorithms are proposed,
the first guarantees progress, the second guarantees
also lockout-freedom. 

There are two  proofs of correctness of the Lehmann and Rabin algorithms, 
one in \cite{Rabin:94:HOARE} and another one, more structured and formal, 
in \cite{Lynch:94:PODC,Segala:95:PhD}.
They both depend in an essential way on the topology. 
In particular, they depend on the fact that one fork can only be shared by two 
philosophers (cfr. Lemma 1 in \cite{Rabin:94:HOARE}, Lemma 7.13 in \cite{Lynch:94:PODC}, 
and Lemma 6.3.14 in \cite{Segala:95:PhD}).
Therefore, a question naturally arises: Would the solutions of Lehmann and Rabin 
still work in the case of more general connection structures? 
The problem is also of practical relevance, since the kind of resource network 
represented by the classic formulation is very restricted.
In this paper we investigate this question and show that the
answer is no: 
In most situations, both the algorithms of Lehmann and Rabin fail. 
We then propose
another solution, still randomized but based on a rather different
idea: we let each philosopher try to establish a partial order on
forks, by assigning a random number to his adjacent forks. In
other words, we use randomization for breaking the initial
symmetry and achieving a situation in which the forks are partially 
ordered. Finally, we propose a variant of the algorithm which ensures
that no philosopher will starve
(lockout-freedom). The algorithms are robust wrt every fair scheduler.

Our motivation for this work comes from the project of providing a distributed 
implementation for the $\pi$-calculus \cite{Milner:92:IC}. 
So far, only the so-called asynchronous subset has been implemented \cite{PIERCE:98:MILNER}. 
In \cite{Palamidessi:97:POPL}, the second author has shown that the full $\pi$-calculus 
is strictly more expressive than its asynchronous subset, and, more in general, that
there is no hope of implementing the $\pi$-calculus with deterministic methods. 
In \cite{Nestmann:97:EXPRESS} Nestmann has shown that the gap in expressive power, 
and the difficulty in the implementation with deterministic methods, is due 
to the mixed (input and output) guarded choice construct of the $\pi$-calculus. 
Such mechanism, however, would be very desirable as 
it provides a powerful programming primitive for solving distributed conflicts. 
Thus, we are considering a randomized implementation. 
We have developed an asynchronous probabilistic 
$\pi$-calculus ($\pi_{pa}$, \cite{Herescu:00:FOSSACS}), 
and we are currently investigating a translation 
from $\pi$ to $\pi_{pa}$ that 
requires solving a resource allocation problem similar to the one of the  
generalized dining philosophers (the resources corresponds the channels 
of the $\pi$-calculus). The restriction to symmetric solutions 
comes from the desire of providing a  {\it compositional} translation.
To our knowledge, there has been only a previous proposal for a symmetric 
and fully distributed implementation of a concurrent
language with guarded choice (\cite{Francez:80:FOCS}), 
but such proposal works only under the assumption of ``good'' schedulers, i.e. a scheduler 
which behaves uniformly through the computation regardless of the actions 
performed by the processes.

\section{The generalized dining philosophers problem}
In this section we introduce a generalization of the dining philosophers
problem. The generalization consists in relaxing
the assumptions about the topology of the system. In the classic 
problem the philosophers and the forks are distributed along a 
ring (table) in alternated position.
On the contrary, we consider arbitrary connection topologies,
and in particular we admit the possibility that a fork is
shared by more than two philosophers. Thus the number of forks and
the number of philosophers is not necessarily the same.
The only constraint we impose on the topology is
that each philosopher is connected (has access) to two distinct forks.
For the rest, the new formulation coincides with the classic one.

\begin{definition}
A generalized dining philosopher system
consists of $n\geq 1$ philosophers and $k\geq 2$ forks.
Unlike the classic case, $n$ and $k$ may be different numbers, and
a fork can be shared by an arbitrary (positive) number of philosophers.
Like in the classic case, every philosopher has access to two forks,
which he will refer to as {\it left} and {\it right}.
Every philosopher can think or eat. When a philosopher wants to eat, he
must pick up the two forks. He can pick up only one fork at the time. He cannot
pick up a fork if his neighbor is already holding it. He cannot eat forever. 
After eating the philosopher releases the two forks and resumes thinking.
\end{definition}

\begin{figure*}[ht]
\begin{center}
\epsfxsize=5in \epsfbox{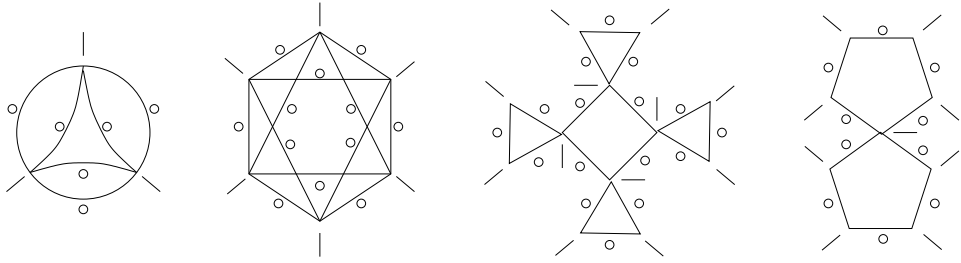}
\end{center}
\caption{Some examples of generalized dining philosophers. From left to right:  
6 philosophers, 3 forks. 12 philosophers, 6 forks. 16 philosophers, 12 forks. 
10 philosophers, 9 forks.}
\label{fig:fig0}
\end{figure*}

Figure \ref{fig:fig0} shows some examples of generalized dining
philosopher systems. Here and in the rest of the paper, we
represent a system as an undirected graph where the nodes are the
forks (represented by sticks in Figure \ref{fig:fig0}), and the
arcs are the philosophers (represented by circles in Figure
\ref{fig:fig0}). Obviously, the forks accessible to a philosopher
are the adjacent nodes. Note that we adopt the more general
definition of graph, which allows the presence of 
more than one arc between two nodes 
(some textbooks use the term {\it multigraph}).

The goal is the same as in the classic problem: to program the philosophers so that
hungry philosophers will eventually eat. Following the standard terminology,
we will say that a solution ensures progress (wrt a set of philosophers) if
it guarantees that, whenever a philosopher of the set is hungry,
then a philosopher of the same set (not necessarily the same philosopher)
will eventually eat\footnote{In \cite{Rabin:94:HOARE} this
property is called {\it deadlock-freedom}.}.
A solution is {\it lockout-free} (wrt a set of philosophers)
if it guarantees that, whenever a philosopher of the set is hungry,
then the same philosopher will eventually eat.

We shall consider only {\it fully distributed} and
{\it symmetric solutions}, namely algorithms where
the only processes are the philosophers, the only shared variables are the forks, 
all philosophers run  identical programs and both the philosophers and the forks
are in the same initial state. We assume that  {\it test-and-set} operations
on the forks are performed atomically. 

A computation consists in an interleaving
of actions performed by the philosophers. Such interleaving is controlled by an
{\it adversary} (or {\it scheduler}). We assume that the adversary has complete
information of the past of the computation, and can decide its next step on the basis
of that information. We consider only {\it fair} adversaries, namely
adversaries that ensure that each philosopher executes infinitely many actions
in each of the possible computations.

We will consider {\it randomized algorithms}, namely algorithms which allow a
philosopher to select randomly between two or more alternatives. The outcome
of the random choice depends on a probability distribution, and it is not
controlled by the adversary. For this reason, even under the same adversary,
different computations may be possible.
This model of computation has been formalized by Lynch and Segala by introducing
the concept of {\it probabilistic automata} \cite{Segala:95:NJC,Segala:95:PhD}.

\section{Limitations of the algorithms of Lehmann and Rabin}
In this section we show that the randomized algorithms of Lehmann and Rabin
presented in \cite{Rabin:94:HOARE} do not work anymore in the general case.

We start by recalling the first algorithm of Lehmann and Rabin, LR1 for short.
Each philosopher runs the code written in Table~\ref{Table LR1}.
\renewcommand{\arraystretch}{1.5}
\begin{table}
\begin{center}$
\begin{array}{ll}
1. &{\it think}; \\
2. &{\it fork} \;:=\; {\it random\_choice}({\it left},{\it right});\\
3. &{\it if} \;{\it isFree}({\it fork}) \; {\it then} \; {\it take}({\it fork}) \; {\it else}\; {\it goto} \;3; \\
4. &{\it if} \;{\it isFree}({\it other}({\it fork})) \\
   & \quad \quad {\it then} \; {\it take}({\it other}({\it fork}))\\
   & \quad \quad {\it else}\; \{ {\it release}({\it fork}); \;{\it goto} \;2\}\\
5. &{\it eat};\\
6. &{\it release}({\it fork}); \; {\it release}({\it other}({\it fork}));\\
7. & goto\; 1;
\end{array}
$\end{center}
\caption{The algorithm LR1.}
\label{Table LR1}
\end{table}
\renewcommand{\arraystretch}{1}

Following standard conventions we assume that the action {\it think} may not terminate, 
while all the other ones are supposed to terminate. 
The test-and-set operations on the forks, in Steps $3$ and $4$, 
are supposed to be executed atomically.
Each outcome (left or right) of the random draw has a positive probability
and the sum of the probabilities is $1$. 
In the classic algorithm the probability is evenly distributed 
($1/2$ for left and $1/2$ for right). However our negative results do not 
depend on this assumption. 

It has been shown in \cite{Rabin:94:HOARE} that, for the classic
dining philosophers, LR1 ensures progress
with probability $1$ under every fair scheduler. 
A more formal proof of this result
can be found in \cite{Lynch:94:PODC,Segala:95:PhD}.

In the  generalized case this result does not hold anymore. Let us
illustrate the situation with an example. We use the following
notation: An empty arrow, associated with a philosopher and
pointing towards a fork, denotes that the philosopher has
committed to that fork (has selected that fork with the random
choice instruction) but he has not taken it yet. A filled arrow
denotes that the philosopher is holding the fork, namely he has
taken the fork and has not released it yet. From now on, we will
represent the nodes (forks) as bullets, instead than as sticks.

For the sake of simplicity, for the moment we relax the fairness requirement.
We will discuss later how to make the example valid also in presence of
fairness.

Consider the system on the leftmost side of Figure \ref{fig:fig0}, and
consider State  $1$ depicted in the figure below.
Clearly, this state is reachable from the initial state
(where all philosophers are at the beginning of the program, i.e. thinking)
with a non-null probability.

\begin{figure}[ht]
\begin{center}
\epsfxsize=3.2in \epsfbox{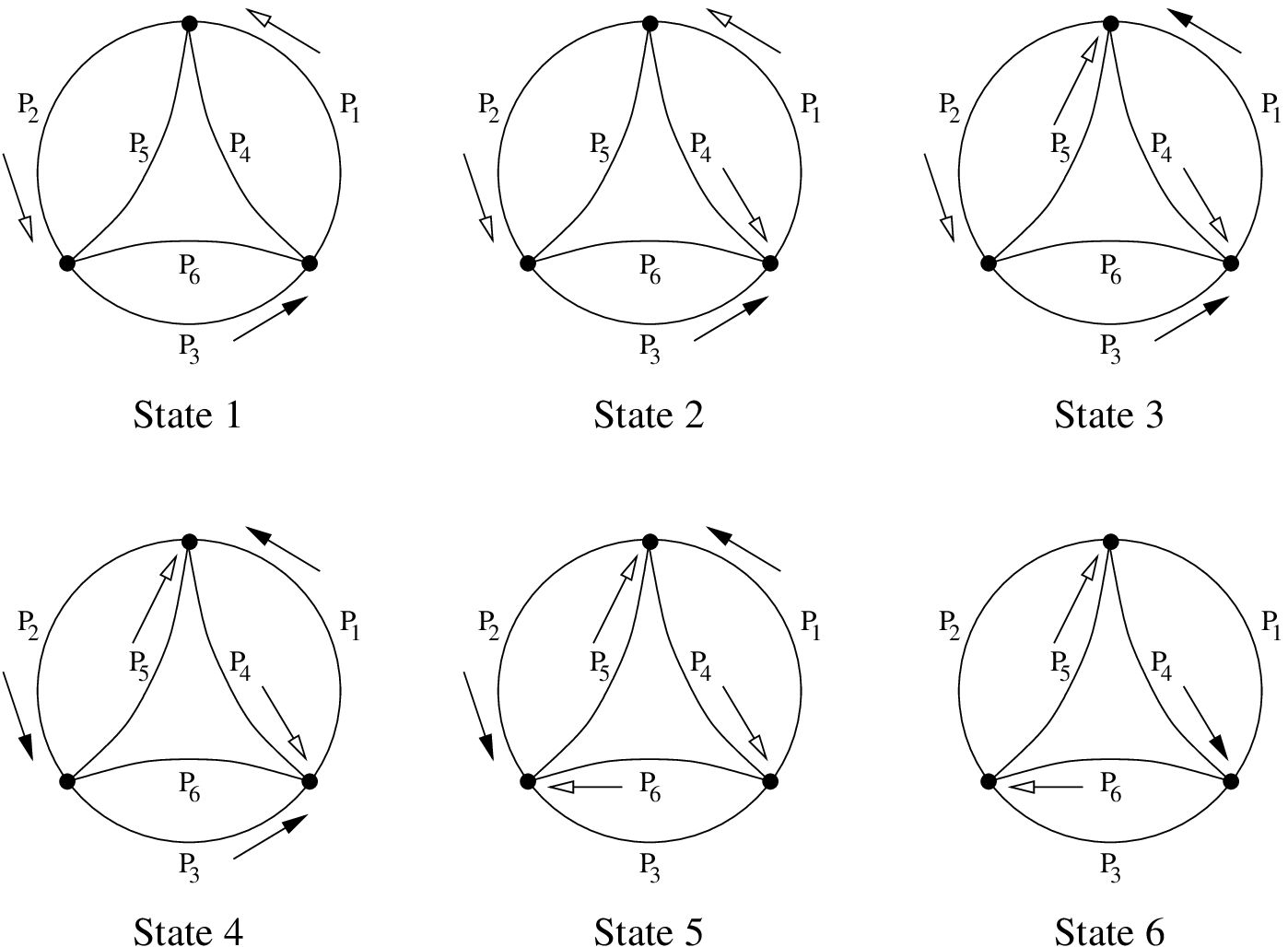}
\end{center}
\end{figure}

The scheduler chooses $P_4$ next. If $P_4$
commits to the free fork, then he will take it and then try to get
the other fork. Since the other fork is taken by $P_3$, $P_4$ has
to release the first fork and draw again. The scheduler keeps selecting $P_4$ until
he commits to the fork taken by $P_3$. When this commitment occurs we are in State $2$.

Next the scheduler selects $P_1$ and $P_1$ takes the fork he had
committed to. Then the scheduler keeps scheduling $P_5$ until $P_5$
commits to the fork taken by $P_1$ (like it was done for $P_4$).
This is State $3$.

Then the scheduler selects $P_2$, and $P_2$ takes the fork he had
committed to. This situation is represented by State $4$.

The scheduler continues with $P_3$. $P_3$ finds his second fork
taken by $P_2$ and therefore releases the fork that he currently
controls. $P_6$ is then scheduled, until it commits to the fork
taken by $P_2$. This is State $5$.

Finally the scheduler runs $P_2$, and $P_2$ will 
have to release his fork, since the other fork is taken by $P_1$. 
Then $P_4$ is selected, and he takes the fork he had committed to.
Then the scheduler selects $P_1$, which will have to release his fork
since the other one is taken by $P_4$. This is State $6$.

Observe now  that State $6$ is isomorphic to State $1$,
in the sense that they differ only for the names of the philosophers.
The scheduler can then go back to State $1$ and then repeat these
steps forever, thus  inducing a computation in which no philosopher is
able to eat. Note that the probability of a computation of this kind is 
$1/4$, which is the probability of reaching a state isomorphic to State $1$ 
already at the first attempt. (We are assuming that the probability of 
picking a particular adjacent fork is evenly distributed between 
left and right, i.e. it's $1/2$. If this is not the case then the 
figure above will be different from $1/4$, but it will still be positive.) 
It's easy to see that, by repeating the 
attempt to reach State $1$ (possibly after some philosopher has eaten), 
the scheduler can {\it eventually} induce 
a cycle like the above one with probability $1$.

Unfortunately the scheduler considered in this example is unfair. In fact, it keeps
selecting one philosopher (for instance $P_4$) until it commits to a taken fork.
If the philosopher chooses forever the free fork, then the resulting computation is unfair.
Although such a computation has probability $0$, according to the definition of fairness,
the scheduler is unfair.

However, it is easy to modify the scheduler so to obtain a fair scheduler
which achieves the same result, namely a no-progress computation with non-null
probability, although smaller than $1/4$.
Consider a variant of the above scheduler which keeps selecting a
``stubborn philosopher''  for a finite number of times only,
but which increases this number at every round. By ``round'' here
we mean the computation fragment which goes from State $1$ to State $6$, and back
to State $1$, as described above. Let $n_k$ be the maximum number of times
which the scheduler is allowed to select the same philosopher during the $k$-th round.
Choose $n_k$ to be big enough so that the probability that the
scheduler actually succeeds to complete the $k$-th round is $1-p^k$ with $p\leq 1/2$.
Consider an infinite computation made of successive successful rounds.
The probability of this computation is greater than or equal to
\[
\frac{1}{4} \; \Pi_{k=1}^\infty (1-p^k).
\]
It is easy to prove by induction that for every $m\geq 1$,
\[
\Pi_{k=1}^m (1-p^k) \geq
1 - p - p^2 + p ^{m+1}
\]
holds. Hence we have 
\[
\Pi_{k=1}^\infty (1-p^k) \geq 1 - p - p^2.
\]
Furthermore, by the assumption $p\leq 1/2$, we have 
\[
1 - p - p^2 \geq 1/4.
\]

\subsection{A general limitation to the first algorithm of Lehmann and Rabin}
We have seen that there is at least one example of graph in which LR1 does not work.
One could hope that this example represents a very special situation, and that 
under some suitable conditions LR1 could still work in more general 
cases than just the standard one.
Unfortunately this is not true: 
It turns out that as soon as we allow one fork of the ring to be shared by an additional  
philosopher, LR1 fails.

In the following, we will call {\it ring} (or {\it cycle}) 
a graph which has $k$ nodes, say $0, 1, \ldots k-1$,
and $k$ arcs connecting the pairs $(0, 1), (1, 2), \ldots, (k-1, 0)$.

\begin{theorem}\label{th1}
Consider a graph $G$ containing a ring subgraph $H$, and such that one of the nodes
of  $H$ has at least three incident arcs (i.e. an additional arc in $G$ besides
the two in $H$). Then it is possible to define a fair scheduler for {\rm LR1} such that
the probability of a computation in which 
the arcs (philosophers) in $H$ make no progress is strictly positive.
\end{theorem}
{\bf Proof (Sketch)}
Figure \ref{fig:proof1} represents the subgraph of $G$ consisting of
{\begin{itemize}
\item the ring $H$
(a hexagon in the figure, but the number of vertices is not important)
with a node $f$ having (at least) three incident arcs,
\item  the arc $P$ in $G$ but not in $H$ which is incident on $f$, and
\item the node $g$ adjacent to $f$ via $P$.
\end{itemize}
It does not matter whether $g$ is a node in $H$ or not.

\begin{figure*}[ht]
\begin{center}
\epsfxsize=3.5in \epsfbox{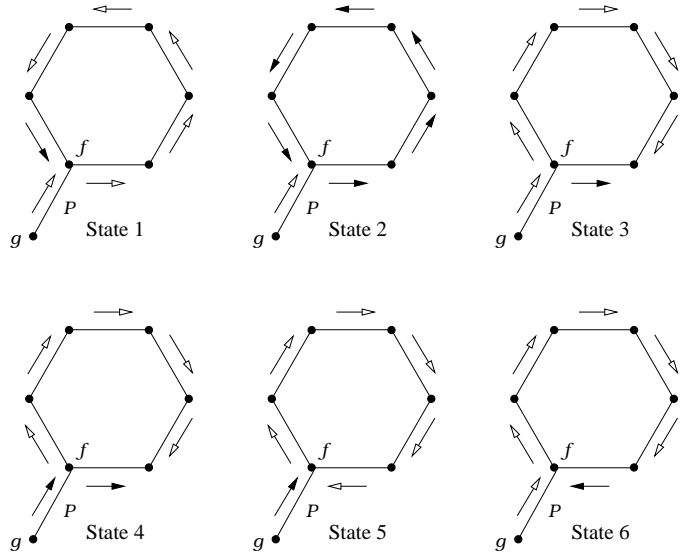}
\end{center}
\caption{A winning scheduling strategy against the algorithm LR1.} \label{fig:proof1}
\end{figure*}

Figure \ref{fig:proof1} shows a possible sequence of states induced by a
scheduler $S$. State $1$ is reachable from the initial state
(where all philosophers are thinking) with a non-null probability.
The scheduler controls the directionality of the arrows
by means of the technique explained in the example in previous section.
The state transitions should be rather clear, except maybe for the last one. That transition
(between State $5$ and State $6$) is achieved by the following sequence of actions:
\begin{itemize}
\item schedule $P$ and let him
eat (this is always possible - the scheduler can always make $g$ free
at the moment $P$ needs it)
\item schedule  the philosopher adjacent to $P$
which is committed to $f$, and let him take $f$
\item keep scheduling $P$ until he commits to $f$.
\end{itemize}

State $6$ is symmetric to State $1$ and we can therefore define an infinite computation,
where no philosopher in $H$ eats, by repeating the actions which bring from State $1$
to State $6$, and then back to State $1$, and so on.

Again, the scheduler $S$ illustrated here is not fair, but we can
obtain a fair scheduler $S'$ which approximates $S$ by letting the
``level of stubbornness'' of $S'$ 
increase at each round, following the technique used in
the example above. \qed

\subsection{The second algorithm of Lehmann and Rabin}\label{LR2}
In this section we consider the second algorithm of Lehmann and Rabin,
presented in \cite{Rabin:94:HOARE} as a lockout-free solution to 
the classic dining philosophers.

We consider here a slight generalization of the original algorithm suitable for the
a generic topology. Hereafter we will refer to it as LR2. 
We assume that each fork is provided with the following data structures:
\begin{itemize}
\item A list of incoming requests $r$, with operations {\it isEmpty}, {\it insert}, and {\it remove}.
Initially the list is empty.
\item A ``guest book'' $g$, namely a list which keeps track of the philosophers
who have used the  fork.
\end{itemize}
The idea is that when a philosopher gets hungry, he inserts his
name {\it id} in the request list of the adjacent 
forks\footnote{We do not need to assume that 
all philosophers have different {\it ids}, 
but simply that those who share a fork
are distinguished form each other. This assumption does not violate 
the symmetry requirement. In fact, the distinction between the 
adjacent philosophers could be stored in the fork and used only 
within the operations on the fork.}. 
After the philosopher has eaten, he removes his name from these lists, and
signs up the  guest books of the forks. Before picking up a fork, a
philosopher must check that there are no other incoming requests
for that fork, or that the other philosophers requesting the fork
have used it after he did. This  condition will be represented, in
the algorithm, by the condition {\it Cond}({\it fork})\footnote{In
the original algorithm the list of incoming request is replaced by
two switches associated to the two adjacent philosophers, and instead of the guest
book there is a simple variable which indicates which one of 
the adjacent philosophers has eaten last.}.

Table~\ref{Table LR2} shows the code run by each philosopher.
\renewcommand{\arraystretch}{1.5}
\begin{table}
\begin{center}
$\begin{array}{ll}
1. &{\it think}; \\
2. &{\it insert}({\it id},{\it left.r}); \; {\it insert}({\it id},{\it right.r}); \\
3. &{\it fork} \;:=\; {\it random\_choice}({\it left},{\it right});\\
4. &{\it if} \;{\it isFree}({\it fork}) \;{\it and} \;{\it Cond}({\it fork})\\
   & \quad \quad {\it then} \; {\it take}({\it fork}) \; \\
   & \quad \quad {\it else}\; {\it goto} \;4; \\
5. &{\it if} \;{\it isFree}({\it other}({\it fork})) \;\\
   & \quad \quad {\it then} \; {\it take}({\it other}({\it fork}))\; \\
   & \quad \quad {\it else}\; \{ {\it release}({\it fork}); \;{\it goto} \;3;\}\\
6. &{\it eat};\\
7. &{\it remove}({\it id},{\it left.r}); \; {\it remove}({\it id},{\it right.r}); \\
8. &{\it insert}({\it id},{\it left.g}); \; {\it insert}({\it id},{\it right.g}); \\
9. &{\it release}({\it fork}); \; {\it release}({\it other}({\it fork}));\\
10. & goto\; 1;
\end{array}$
\end{center}
\caption{The algorithm LR2.}
\label{Table LR2}
\end{table}
\renewcommand{\arraystretch}{1}

The negative result expressed in Theorem \ref{th1} does not hold for LR2.
In fact, once $P$ has eaten, he cannot take Fork $f$ before the neighbor
has eaten as well. However the class of graphs in which LR2 does not work is still fairly general:

\begin{theorem}
Consider a graph $G$ containing a ring subgraph $H$, and such
that two of the nodes in  $H$ are connected at least by three
different paths (i.e. an additional path $P$ in $G$ besides the two in
$H$). Then it is possible to define a fair scheduler for {\rm
LR2} such that the probability of a computation in which the arcs (philosophers) 
of $H$ and $P$ make no
progress is strictly positive.
\end{theorem}
{\bf Proof (Sketch)}
The proof is illustrated in Figure \ref{fig:proof2}, which shows the part of
$G$ containing the ring $H$ and the additional path between two nodes of $H$.

\begin{figure*}[ht]
\begin{center}
\epsfxsize=3.5in\epsfbox{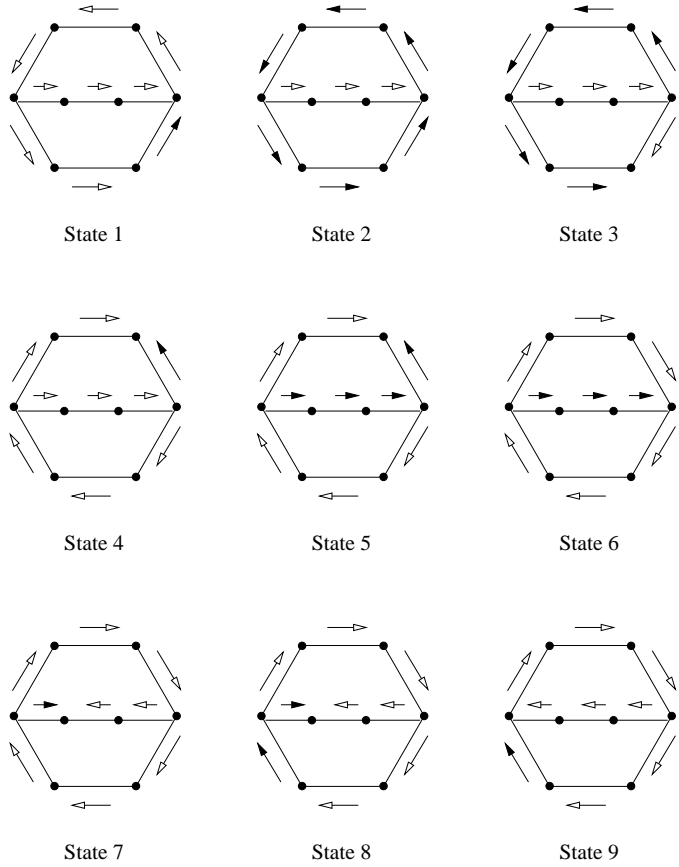}
\end{center}
\caption{A winning scheduling strategy against the algorithms LR1 and LR2.} \label{fig:proof2}
\end{figure*}

Like before, the computation illustrated in the figure is 
induced by an unfair scheduler $S$, but by following the 
usual technique we can define a fair scheduler $S'$ which 
approximates $S$ and which achieves the same result. 
Note that none of the philosophers
in $H$ and in the additional path  ever gets to eat, 
hence the modification of LR2 wrt LR1, namely the test 
{\it Cond}({\it fork}), is useless: {\it fork.g} remains forever empty.
\qed

\section{A deadlock-free solution}
In this section we propose a symmetric and fully distributed
solution to the generalized dining philosophers and show that it makes progress with
probability $1$.

Our algorithm works as follows. Each fork has associated
a field {\it nr} which contains an integer number ranging in the interval
$[0,m]$, with $m\geq k$, where $k$ is the total number of forks in the system.
Initially  {\it nr} is $0$ for all the forks.
Each philosopher can change the {\it nr} value of a fork when he gets hold of it,
and he tries to make sure that  the {\it nr} values of  its adjacent
forks become and are maintained different. In order to ensure that this situation
will be eventually achieved, each new {\it nr} value is chosen
{\it randomly}. Note that this random choice  is necessary to break the symmetry,
otherwise, in presence of a ring, a malicious scheduler
could induce a situation where one philosopher changes one fork, then his neighbor
changes the other fork to the same value, and so on, for all the forks in the ring.

Our algorithm is similar to LR1, except that the choice of the first fork is
done by picking the one with the highest {\it nr} value (if they are different),
instead than randomly. The other difference is that, as explained before, 
the philosopher
may change the {\it nr} value of a fork when it finds that it is equal to the 
{\it nr} value
of the other fork. This is done by calling ${\it random}[1,m]$, which 
returns a natural number in the interval $[1,m]$, selected probabilistically. 
We assume for simplicity that 
the probability of the 
outcome is evenly distributed among the numbers in the interval.
The algorithm is illustrated in Table~\ref{Table GDP1}. We will refer to it  as GDP1.

\renewcommand{\arraystretch}{1.5}
\begin{table}
$$
\begin{array}{ll}
1. &{\it think}; \\
2. &{\it if} \; {\it left.nr} > {\it right.nr} \\
   & \quad \quad  {\it then} \; {\it fork} \mbox{ := } {\it left} \\
   & \quad \quad {\it else} \; {\it fork} \mbox{ := } {\it right};  \\
3. &{\it if} \;{\it isFree}({\it fork}) \; {\it then} \; {\it take}({\it fork}) \; {\it else}\; {\it goto} \;3; \\
4. &{\it if} \;{\it fork.nr} = {\it other}({\it fork}){\it .nr} \\
   & \quad \quad  {\it then} \; {\it fork} \mbox{ := } {\it random}[1,m];  \\
5. &{\it if} \;{\it isFree}({\it other}({\it fork}))\\
   & \quad \quad {\it then} \; {\it take}({\it other}({\it fork}))\\
   & \quad \quad {\it else}\; \{ {\it release}({\it fork}); \;{\it goto} \;2\}\\
6. &{\it eat};\\
7. &{\it release}({\it fork}); \; {\it release}({\it other}({\it fork}));\\
8. & goto\; 1;
\end{array}
$$
\caption{The algorithm GDP1.}
\label{Table GDP1}
\end{table}
\renewcommand{\arraystretch}{1}

We prove now that GDP1 makes progress, under every fair scheduler,
with probability $1$.
The proof is formalized in terms of the {\it progress} and {\it unless} statements introduced in
\cite{Lynch:94:PODC,Segala:95:PhD}. A progress statement is denoted by \ $S
\catarrow{p}{A}  S'$, where $S$ and $S'$ are sets of states, $p$
is a probability, and $A$ is a class of adversaries. Its meaning
is that starting from any state in $S$, under any adversary in
$A$, a state in $S'$ is reached with probability at least $p$.
An unless statement is of the form $S\; {\it unless}\; S'$ and means that, if the system 
is in one of the states of $S$, then it remains in $S$ (possibly moving through different 
states of $S$) 
until it reaches a state in $S'$.

Let us define $T$ to be the set of states in which some philosopher
tries to eat ({\it trying section}, steps 2 through 5), and  $E$ to be the set of states in
which some philosopher is eating. We will show that \ $T \catarrow{1}{F}  E$,
where $F$ is the class of all fair adversaries.
The following properties of progress statement, proved in  
\cite{Lynch:94:PODC,Segala:95:PhD},
will be useful for our purposes.

\begin{lemma}[Concatenation]\label{lemma1}\ \\
If \  $S \catarrow{p_1}{A}  S'$ \ and \ $S' 
\catarrow{p'}{A}  S''$, then \  $S  \catarrow{p p'}{A} S''$.
\end{lemma}

\begin{lemma}[Union]\label{lemma2}\ \\
If \  $S_1  \catarrow{p_1}{A}  S'_1$ \ and \  $S_2 
\catarrow{p_2}{A}  S'_2$, then \  $S_1 \cup S_2  \catarrow{p}{A}  S'_1 \cup S'_2$
\ with \ $p= {\it min}\{p_1,p_2\}$.
\end{lemma}

\begin{lemma}[Persistence wins]\label{lemma3}\ \\
If \ $S  \catarrow{p}{F}  S'$ \ with $p > 0$, and \ $S \;{\it unless} \;S'$,
then \ $S  \catarrow{1}{F}  S'$.
\end{lemma}

We are now ready to prove the correctness of GDP1. 

\begin{theorem}\label{deadlock-freedom} \ 
$T \catarrow{1}{F}  E$.
\end{theorem}
{\bf Proof}
\ Let us denote by  $C_1$ the set of states in which there
is one cycle in the graph where all adjacent forks have different
numbers. $C_2$ is the set of states in which there are two cycles in the
graph where all adjacent forks have different numbers, and so on.

We have the following progress statements:
\begin{itemize}
\item
$T \catarrow{p}{F} (T \cap C_1) \cup E$, with $p\geq {m!}/{m^k  (m-k)!}$.
In fact, one of the trying philosophers, say $P_1$, will find the first fork free
and will pick it up. Then, either he will find also his second fork free, and therefore
will eat, or it will find the second fork taken by another philosopher, say $P_2$.
Again, either $P_2$ will eat, or will find his second fork taken by another philosopher,
say $P_3$, etc. Since the number of philosophers is finite, we will end up either with one
philosopher eating, or with a ring of forks all picked up as first forks at least once.
Since each philosopher changes the {\it nr} value  of the  first fork if this value is equal to
that of the other fork, the adjacent forks of this ring will get all different values with
probability $p$ not smaller than ${m!}/{m^k  (m-k)!}$ (this is the probability that,
if we assign randomly values in the range $[1,m]$ to the nodes of a complete graph of cardinality $k$,
all the nodes get a different value). Note that, by the assumption $m\geq k$, we have $p>0$.
\item $T \cap C_1 \catarrow{p}{F}  (T \cap C_2) \cup E$. Similar to previous point.
\item ...
\item $T \cap C_{h-1} \catarrow{p}{F}  (T \cap C_h) \cup E$. Similar to previous point.
\item $T \cap C_{h} \catarrow{1}{F}   E$. When all possible cycles in the graph
have adjacent nodes with different {\it nr} values, then the algorithm works like a
hierarchical resource allocation algorithm based on a partial ordering: Let $P$
be the first philosopher who is holding the first fork, and such that
the {\it nr} value of the other fork $f$ is the smallest of all the forks adjacent to $f$.
Then either $P$ or one of his neighbors will eat.
\end{itemize}
From the above statements, and by using Lemma \ref{lemma1}, the
obvious fact that \ $E \catarrow{1}{F}   E$,  and Lemma
\ref{lemma2}, we derive
\[
T \catarrow{p^h}{F}   E.
\]
On the other hand, it's clear that, since philosophers keep trying until they eat, we have also
\[
T \; {\it unless} \; E.
\]
Therefore, by applying Lemma \ref{lemma3}, we conclude $T \catarrow{1}{F}  E$.
\qed

Note that GDP1 does not guarantee that we will reach, with probability 1,
a situation where all adjacent forks will
have a different {\it nr}. Not even if all philosophers 
are in the trying section infinitely often. 
This is because some philosophers may never succeed to pick up a fork,
for instance because they are always scheduled when their neighbors are eating.

\section{A lockout-free solution}

The algorithm GDP1 presented in the previous section is not
lockout-free. In fact, consider two adjacent philosophers, $P_1$
and $P_2$, which share a fork $f$ with a $\it nr$ value which is
smaller than the value of the other fork $g$ of $P_1$. Then $P_1$
will keep selecting $g$ as first fork, and the scheduler could
keep scheduling the attempt of  $P_1$ to pick the second fork,
$f$, only when $f$ is held by $P_2$.

We now propose a lockout-free  variant of GDP1. The idea is to associate to each fork
a list of incoming requests $r$, and a ``guest book'' $g$, like it was done in Section~\ref{LR2}.
The test {\it Cond}({\it fork}) is defined in the same way as in Section~\ref{LR2}.
The new algorithm, that we will call GDP2, is shown in Table~\ref{Table GDP2}.

\renewcommand{\arraystretch}{1.5}
\begin{table}
$$
\begin{array}{ll}
1. &{\it think}; \\
2. &{\it insert}({\it id},{\it left.r}); \; {\it insert}({\it id},{\it right.r}); \\
3. &{\it if} \; {\it left.nr} > {\it right.nr} \\
   & \quad \quad  {\it then} \; {\it fork} \mbox{ := } {\it left} \\
   & \quad \quad  {\it else} \; {\it fork} \mbox{ := } {\it right};  \\
4. &{\it if} \;{\it isFree}({\it fork}) \; {\it then} \; {\it take}({\it fork}) \; {\it else}\; {\it goto} \;4; \\
5. &{\it if} \;{\it fork.nr} = {\it other}({\it fork}){\it .nr} \\
   & \quad \quad  {\it then} \; {\it fork} \mbox{ := } {\it random}[1,m];  \\
6. &{\it if} \;{\it isFree}({\it other}({\it fork})) \\
   & \quad \quad  {\it then} \; {\it take}({\it other}({\it fork}))\\
   & \quad \quad {\it else}\; \{ {\it release}({\it fork}); \;{\it goto} \;3\}\\
7. &{\it eat};\\
8. &{\it remove}({\it id},{\it left.r}); \; {\it remove}({\it id},{\it right.r}); \\
9. &{\it insert}({\it id},{\it left.g}); \; {\it insert}({\it id},{\it right.g}); \\
10. &{\it release}({\it fork}); \; {\it release}({\it other}({\it fork}));\\
11. & goto\; 1;
\end{array}
$$
\caption{The algorithm GDP2.}
\label{Table GDP2}
\end{table}
\renewcommand{\arraystretch}{1}

We show now that GDP2 is lockout-free. 
In the following, $T_i$ will represent the set of states 
in which the philosopher 
$P_i$ is trying to eat, and
$E_i$ the situation in which the philosopher 
$P_i$ is eating.

\begin{theorem} \ 
$T_i  \catarrow{1}{F} E_i$.
\end{theorem}
{\bf Proof} \ 
Let us denote by $C_{i,r}$ is the set of states in which there 
are $r$ cycles containing the arc $P_i$, 
and where all adjacent forks have different numbers. Furthermore, let us use 
$W_{i,s}$ to represent the set of states in which there 
are $s$ philosophers connected to $P_i$ which have already eaten and 
can't eat until all their adjacent philosophers (and ultimately $P_i$)
have eaten as well. 

The proof is similar to the one of Theorem \ref{deadlock-freedom}. 
The invariant in this case is
\[
T_i\cap C_{i,r} \cap W_{i,s}\  \catarrow{p}{F}\  \begin{array}{l}
                                             (T_i\cap C_{i,r+1} \cap W_{i,s})\\ 
                                             \cup \\
                                             (T_i\cap C_{i,r} \cap W_{i,s+1}) \cup E_i
                                             \end{array}
\]
where $p$ has the same lower bound as in the proof of Theorem \ref{deadlock-freedom}.
Furthermore, if $h$ is the total number of cycles containing $P_i$, and $m$ is the 
total number of philosophers connected to $P_i$, we have 
\[
T_i\cap C_{i,h} \cap W_{i,s} \catarrow{p}{F} 
(T_i\cap C_{i,h} \cap W_{i,s+1}) \cup E_i,
\]
\[
T_i\cap C_{i,r} \cap W_{i,m} \catarrow{p}{F} (T_i\cap C_{i,r+1} \cap W_{i,m}) \cup E_i,
\]
and
\[
T_i\cap C_{i,h} \cap W_{i,m} \catarrow{1}{F} E_i.
\]
Hence, by Lemma \ref{lemma1} and \ref{lemma2} we derive
\[
T_i\catarrow{p^{h+m}}{F} E_i.
\]
Since \ $T_i \;{\it unless} \; E_i$, by Lemma \ref{lemma3} we conclude.
\qed

\section{Conclusion and future work}
We have shown that the randomized  algorithms of Lehmann and Rabin for the
symmetric  and fully distributed dining philosophers problem
 do not  work anymore when we relax
the condition that the topology is a simple ring.
We have then proposed randomized solutions ensuring progress and lockout-freedom
for the general case of an arbitrary connection graph.

In this paper we have focused on the existence of a solution, and
we have not address any efficiency issue. Clearly, efficiency is an
important attribute for an algorithm. The evaluation of the complexity 
of our algorithms, and possibly the study of more efficient
variants, are open topics for future research.

Another open problem that seems worth exploring is 
the symmetric and fully distributed solution
in the even more general case of hypergraphs-like connection
structures, in which a philosopher may need more
than two forks to eat.

This work is part of a project which aims at providing a fully 
distributed implementation 
of the $\pi$-calculus. The algorithm presented here will serve for 
solving the conflicts associated to the competition for channels arising in presence of 
guarded-choice commands. 

\section{Acknowledgements}
The authors would like to thank Dale Miller and the anonymous referees of PODC 2001 
for their helpful comments.

\bibliographystyle{abbrv}
\bibliography{/home/users1/catuscia/BIBLIOGRAFIE/biblio_cat}
\end{document}